# Significant Impact of Quantum and Anharmonic Effects on the Structural Stability and Superconductivity of NbH$_3$ at High Pressures

Pugeng Hou[1†], Yao Ma[2†], Hui Xie[3], Mingqi Li[2], Yongmao Cai[1], Yuhua Shen[1], Xuewu Wang[1]*, Mi Pang[2]*

1 College of Science, Northeast Electric Power University, Changchun Road 169, 132012, Jilin, P. R. China
2 Department of Applied Physics, School of Sciences, Xi'an University of Technology, Xi'an 710048, P. R. China
3 College of Physics and Electronic Engineering, Hebei Minzu Normal University, Chengde, 067000, P. R. China

First-principles calculations combined with the stochastic self-consistent harmonic approximation reveal significant effects of the quantum ionic fluctuations and lattice anharmonicity on the dynamical stability of NbH$_3$ under high pressures. Previous theoretical predictions, which ignored ionic fluctuations and relied on the harmonic approximation, suggested that the $I4_3d$ phase is the most thermodynamically favorable structure between 33 and 400 GPa, with the $Fm\bar{3}m$ phase considered thermodynamically metastable. However, recent experiments at 187 GPa identified the $Fm\bar{3}m$ phase, conflicting with the prediction. In contrast, the present study indicates that the $Fm\bar{3}m$ phase remains dynamically stable down to at least 145 GPa, approximately 145 GPa lower than harmonic estimates, while the $I4_3d$ phase is dynamically unstable at 187 GPa, consistent with the experimental findings. Furthermore, systematic calculations are performed on the structural, vibrational and superconducting properties of $Fm\bar{3}m$ NbH$_3$ under pressures ranging from 100 to 300 GPa, revealing dramatic modifications due to the quantum and anharmonic effects. The calculated superconducting critical temperature ($T_c$) from the McMillan equation for $Fm\bar{3}m$ NbH$_3$ at 187 GPa is 44 K, with $\mu^*$ set at 0.15, close to the measured value. These findings highlight the crucial role of quantum anharmonic effects in stabilizing the $Fm\bar{3}m$ phase.

[†] **Pugeng Hou** and **Yao Ma** contribute equally to this work.
*Corresponding author: wangxw@neepu.edu.cn and pangmi@xaut.edu.cn

# I. INTRODUCTION

Driven by the pursuit of metallic and superconducting hydrogen at extreme pressures[1], the integration of first-principles structural predictions with electron-phonon interaction calculations has, in recent years, yielded the prediction of numerous superconducting hydrides featuring elevated superconducting critical temperatures ($T_c$)[2-26]. Superconducting transitions exceeding 200 K have been experimentally observed in sulfur[27] and lanthanum[28-29] super-hydrides at pressures exceeding 100 GPa. Both compounds have been successfully synthesized[27-28] experimentally, following their theoretical anticipation[2-3]. The capacity of density-functional theory (DFT)-based first-principles calculations to guide experimental pursuits has been well-established[2–8].

Most of the structural and superconducting predictions performed so far rely on the classical picture, where the atoms with positions denoted by R, oscillate around their equilibrium positions $R_0$, which are assumed to be the minimum of the Born–Oppenheimer potential V(R). V(R) is usually expanded around $R_0$ up to the second order, known as the harmonic approximation. Let a and b be indices that label both an atom in the crystal and a Cartesian direction, the phonon frequencies are determined from the force constant matrix $[\frac{\partial^2 V(R)}{\partial R_a \partial R_b}]_{R_0}$. In this picture, the quantum nature of ionic fluctuations and anharmonicity of the potential are ignored. Calculations based on the harmonic approximation have achieved significant success in the past. However, research in recent years has led to a developing consensus that quantum ionic fluctuations, coupled with the resulting anharmonicity, can significantly influence the crystal structure and phonon spectrum of hydrogen-rich compounds, thereby substantially altering the predicted critical temperature ($T_c$) due to hydrogen's lightness[37-48]. For instance, in palladium[30,31], platinum[31], and aluminium[41] hydrides, anharmonicity hinders H-character optical modes and markedly suppresses superconductivity. In stark contrast, the molecular *Cmca* phase of hydrogen exhibits a different behavior: anharmonicity effectively doubles the $T_c$ from around 100 K to values exceeding 200 K[34], achieved through an increase in the intramolecular distance of approximately 6%. In LaH$_{10}$, it has been suggested that quantum effects stabilize the crystal structure, enabling strong electron-phonon coupling that would otherwise be dynamically unstable[39]. Similarly, in ScH$_6$, the enhancement of the critical temperature by approximately 15% is attributed to the stretching of H2 molecular-like units by about 5%, with the symmetry preserved in the *P*6$_3$/*mmc* space group[42]. The precise manner in which quantum anharmonic effects modulate crystal structure and superconductivity remains an area of active research and is not yet fully understood.

To date, very few predictions of stable metal poly-hydrides have been thoroughly examined or confirmed due to current experimental limitations. Additionally, there have been fewer theoretical studies that produce results consistent with experiments conducted under high pressure. Recently, the experimental synthesis and metallization of NbH$_3$ have been achieved at approximately 187 GPa[49], a pressure significantly exceeding the 100 GPa threshold reported in previous studies that confirmed the existence of several NbH$_{2.5}$ structures[50]. However, the $Fm\bar{3}m$ phase, which has been identified by the experiments[49], appears to be thermodynamically metastable according to previous classical calculations[51]. In contrast, the $I4_3d$ phase, determined as the most thermodynamically and dynamically stable structure by earlier predictions[51], was not detected in the experiments. Additionally, while the critical temperature ($T_c$) of *I4/mmm*-NbH$_4$ is predicted to surpass 49 K[52] at 300 GPa, it fails to account for the $T_c$ measured at 187 GPa[49], as the dynamically stable pressure for this structure is estimated to be above 287 GPa[52]. These inconsistencies highlight the need for a more in-depth analysis, independent of perturbative methods, to ascertain whether anharmonic effects are responsible for the absence of the $I4_3d$ phase and the emergence of the $Fm\bar{3}m$ phase in experimental findings.

This study presents a first-principles investigation into the effects of quantum ionic fluctuations and anharmonicity on the anticipated high-temperature superconducting properties of $Fm\bar{3}m$ NbH$_3$ under high pressure, utilizing the self-consistent harmonic approximation method to account for the quantum and anharmonic effects beyond perturbation theory. We find that the $Fm\bar{3}m$ phase remains dynamically stable down to at least 145 GPa—approximately 145 GPa lower than that predicted by harmonic approximation. In contrast, the $I4_3d$ phase, considered as the most enthalpically favorable structure based on harmonic calculations, is found to be dynamically unstable in our anharmonic analysis. Additionally, the electron-phonon coupling constants ($\lambda$) for $Fm\bar{3}m$ NbH$_3$ decrease under pressure, from 2.14 at 167 GPa to 1.65 at 227 GPa. Notably, this reduction in $\lambda$ does not result in a rapid decline in the critical temperature ($T_c$). The anharmonic $T_c$ calculated for $Fm\bar{3}m$ NbH$_3$ using the McMillan equation aligns well with experimental observations. Our results underscore the key role of quantum anharmonicity in determining the crystal structure of NbH$_3$ at high pressures. This paper is organized as follows: Section II outlines the theoretical framework of our anharmonic ab initio calculations, Section III details the computational methodology used, Section IV presents the results and discussion, and Section V summarizes the main conclusions of this work.

## II. METHODOLOGY

The influence of quantum ionic fluctuations and anharmonicity is evaluated utilizing the stochastic self-consistent harmonic approximation (SSCHA) code[43], which builds upon the theoretical groundwork laid out earlier[30, 31, 36]. This section provides a brief summary of the SSCHA methodology, along with the theoretical framework employed to calculate the superconducting critical temperature, incorporating the effects of anharmonicity.

### A. The stochastic self-consistent harmonic approximation

Without approximating the Born–Oppenheimer potential V(R), the SSCHA rigorously incorporates effects of the quantum fluctuations of ions through a quantum variational technique, aiming at minimizing the system's free energy, which can be computed with a trial density matix $\tilde{\rho}_{R,\Phi}$ as

$$\mathcal{F}[\tilde{\rho}_{R,\Phi}] = <K + V(\mathbf{R})>_{\tilde{\rho}_{R,\Phi}} - TS[\tilde{\rho}_{R,\Phi}]. \tag{1}$$

In the above equation, $K$ and $V(\mathbf{R})$ represent the ionic kinetic energy and Born-Oppenheimer potential respectively. $T$ and $S[\tilde{\rho}_{R,\Phi}]$ denote the temperature and entropy, respectively. The density matrix is parameterized by centroid positions $\mathbf{R}$, determining the average ionic positions, and auxiliary force constants $\Phi$, which relate to the broadening of the ionic wave functions around $\mathbf{R}$. By minimizing $\mathcal{F}[\tilde{\rho}_{R,\Phi}]$ with respect to $\mathbf{R}$ and $\Phi$, a robust variational approximation of the free energy can be achieved without resorting to approximations in the Born-Oppenheimer potential. At the minimum, the resulting positions $\mathbf{R}_{eq}$ establish the average ionic positions, while the auxiliary force constants $\Phi_{eq}$ are associated with the fluctuations around these positions. In contrast, in the classical harmonic approximation, the equilibrium positions $R_0$ are determined by minimizing $V(\mathbf{R})$, which typically differ from $\mathbf{R}_{eq}$ because the ionic kinetic energy is solely ignored. The SSCHA code is capable of optimizing the crystal structure, including lattice degrees of freedom, and fully incorporating ionic quantum effects and anharmonicity at any desired pressure level.

In the static limits[43-45], phonon frequencies are determined from eigenvalues of the mass rescaled second order derivatives of the free energy taken at $\mathbf{R}_{eq}$.

$$\mathbf{D}_{ab}^{(F)} = \frac{1}{\sqrt{M_a M_b}} \frac{\partial^2 \mathcal{F}}{\partial R_a \partial R_b} |_{R_{eq}}, \tag{2}$$

Here, a and b are combined indices for the atoms and their Cartesian coordinates, and $M_a$ represents the mass of atom "a". The dynamical matrix $\mathbf{D}^{(F)}$, also called the free energy Hessian, serves as the quantum anharmonic counterpart to the classical harmonic dynamical matrix, which, instead, is derived from the Hessian of the Born-Oppenheimer potential evaluated at $R_0$:

$$\mathbf{D}_{ab}^{(h)} = \frac{1}{\sqrt{M_a M_b}} \frac{\partial^2 V(R)}{\partial R_a \partial R_b} |_{R_0}. \tag{3}$$

It is worth noting that a negative eigenvalue of $\mathbf{D}^{(F)}$ signals the structural instability in the free energy landscape, including quantum effects and anharmonicity. Similarly, a negative eigenvalue of $\mathbf{D}^{(h)}$ signals classical instability when ionic quantum effects are neglected.

In addition to optimizing the internal ionic positions through the minimization of $\mathcal{F}[\tilde{\rho}_{R,\Phi}]$ with respect to the centroid positions, SSCHA can also relax the lattice cell parameters, including quantum effects and anharmonicity. This is achieved by computing the stress tensor from the derivative of the SSCHA free energy with respect to the components of the strain tensor ϵ:

$$P_{\alpha\beta} = -\frac{1}{\Omega_V} \frac{\partial \mathcal{F}}{\partial \epsilon_{\alpha\beta}} |_{\epsilon=0}, \tag{4}$$

where $\Omega_V$ represents the simulation box volume[43], and α and β denote Cartesian indices. This expression can incorporate the extra pressure induced by ionic fluctuations in addition to the classical harmonic pressure, which is computed by substituting the SSCHA free energy with $V(\mathbf{R})$ in Eq. (4).

## B. Calculation of the superconducting transition temperature

We evaluated the $T_c$ with the McMillan equation[53] and Allen-Dynes modified McMillan equation[54],

$$T_c = \frac{\omega_{log}}{1.2} \exp\left[-\frac{1.04(1+\lambda)}{\lambda-\mu^*(1+0.62\lambda)}\right], \quad (5)$$

$$T_c = \frac{f_1 f_2 \omega_{log}}{1.2} \exp\left[-\frac{1.04(1+\lambda)}{\lambda-\mu^*(1+0.62\lambda)}\right], \quad (6)$$

where $\lambda$ represents the electron-phonon coupling constant, and $\mu*$ is the Coulomb pseudopotential[54]. These equations has yielded $T_c$ values that agree well with experimental results in super-hydrides despite its simplicity[33,39]. $\lambda$ is defined as the first reciprocal moment of the electron-phonon Eliashberg function $\alpha^2 F(\omega)$,

$$\lambda = 2 \int_0^\infty \frac{\alpha^2 F(\omega)}{\omega} d\omega. \quad (7)$$

The other parameters in Eqs. (5) and (6) are calculated as follows:

$$\omega_{log} = \exp\left[\frac{2}{\lambda}\int_0^\infty \frac{d\omega}{\omega} \alpha^2 F(\omega) \ln \omega\right], \quad (8)$$

$$f_1 = \sqrt[3]{\left[1 + \left(\frac{\lambda}{\Lambda_1}\right)^{\frac{3}{2}}\right]}, \quad (9)$$

$$f_2 = 1 + \frac{(\frac{\bar{\omega}_2}{\omega_{log}}-1)\lambda^2}{\lambda^2 + \Lambda_2^2}. \quad (10)$$

$\Lambda_1$, $\Lambda_2$ and $\bar{\omega}_2$ are given by

$$\Lambda_1 = 2.46(1 + 3.8\mu^*), \quad (11)$$

$$\Lambda_2 = 1.82(1 + 6.3\mu^*)\frac{\bar{\omega}_2}{\omega_{log}}, \quad (12)$$

$$\bar{\omega}_2 = \sqrt{\frac{2}{\lambda}\int_0^\infty \alpha^2 F(\omega)\omega \, d\omega}. \quad (13)$$

We calculate the Eliashberg function as

$$\alpha^2 F(\omega) = \frac{1}{2N(0)N_q N_k} \sum_{knm,\mu q,\bar{a}\bar{b}} \frac{\epsilon_\mu^{\bar{a}}(q)\epsilon_\mu^{\bar{b}}(q)^*}{\omega_\mu(q)\sqrt{M_{\bar{a}}M_{\bar{b}}}} \times d_{kn,k+qm}^{\bar{a}} d_{kn,k+qm}^{\bar{b}*} \delta(\varepsilon_{kn})\delta(\varepsilon_{k+qm})\delta(\omega - \omega_\mu(q)). \quad (14)$$

In the equation above, $d_{kn,k+qm}^{\bar{a}} = <kn|\delta V_{KS}/\delta R^{\bar{a}}(q)|k+qm>$, where $|kn>$ represents a Kohn-Sham state with energy $\varepsilon_{kn}$ measured from the Fermi level, $V_{KS}$ denotes the Kohn-Sham potential, and $R^{\bar{a}}(q)$ stands for the Fourier-transformed displacement of atom ā; The combined atom and Cartesian indices with a bar (ā) only run for atoms within the unit cell. $N_k$ and $N_q$ are the number of electron and phonon momentum points utilized for BZ sampling; $N(0)$ represents the density of states at the Fermi level, while $\omega_\mu(q)$ and $\epsilon_\mu^{\bar{a}}(q)$ represent phonon frequencies and the polarization vectors, respectively. In this study, the Eliashberg function is computed both at the harmonic and anharmonic levels, by substituting into Eqs. (14) the harmonic phonon frequencies and polarization vectors obtained by diagonalizing $\mathbf{D}^{(h)}$ or their anharmonic equivalents from diagonalizing $\mathbf{D}^{(F)}$. It is important to note that the derivatives of the Kohn–Sham potential used in the electron–phonon matrix elements are evaluated at different positions in classical harmonic and quantum anharmonic calculations: in the former, they are computed at the positions $R_0$ minimizing $V(\mathbf{R})$, while in the latter, they are determined at the positions $\mathbf{R}_{eq}$ that minimize instead $\mathcal{F}[\tilde{\rho}_{\mathbf{R},\Phi}]$. For comparison, $T_c$ is also determined from solving the isotropic Migdal-Eliashberg equations once $\alpha^2 F(\omega)$ is obtained[55].

## III. COMPUTATIONAL DETAILS

The ab initio calculations were performed using the QUANTUM ESPRESSO (QE) package[56], employing ultrasoft pseudopotentials[57] with the Perdew-Burke-Ernzerhof (PBE) parametrization[58] of the exchange correlation potential. The plane-wave basis cutoff was set to 80 Ry and 800 Ry for the density. First BZ integrations were performed on a $16 \times 16 \times 16$ Monkhorst-Pack mesh[59] using a smearing parameter of 0.01 Ry, $8 \times 8 \times 8$ ($Fm\bar{3}m$) and $4 \times 4 \times 4$ ($I4_3d$) q-point mesh. SSCHA minimization[43] requires the calculation of energies, forces, and stress tensors in supercells. These calculations were conducted within DFT[60] at the PBE level using Quantum ESPRESSO, employing the same pseudopotentials. We performed the calculations in a $2 \times 2 \times 2$ supercell containing 32 atoms for $Fm\bar{3}m$ and 256 atoms for $I4_3d$ NbH$_3$, resulting in dynamical matrices on a commensurate $2 \times 2 \times 2$ grid. A 60 Ry energy cutoff and a $6 \times 6 \times 6$ k-point mesh ($Fm\bar{3}m$) and a 50 Ry energy cutoff and a $4 \times 4 \times 4$ k-point mesh ($I4_3d$) for Brillouin zone (BZ) integrations were sufficient to converge the SSCHA gradient in the supercell. The SSCHA calculations were performed at 0 K. Following each minimization iteration, we augmented the population with a greater number of individuals N using the minimized trial density matrix until convergence was achieved. Two criteria were employed to terminate the minimization loops: firstly, a Kong-Liu ratio, which assesses the effective sample size and should attain a value of 0.5, and secondly, a ratio of less than $10^{-9}$ between the free energy gradient and its stochastic error. The difference between the harmonic and anharmonic dynamical matrices on the $2 \times 2 \times 2$ grid was interpolated to $8 \times 8 \times 8$ grid ($Fm\bar{3}m$) and $4 \times 4 \times 4$ grid ($I4_3d$). By summing the harmonic dynamical matrices in this fine grid to the result, the anharmonic dynamical matrices on the $8 \times 8 \times 8$ grid ($Fm\bar{3}m$) and $4 \times 4 \times 4$ grid ($I4_3d$) were obtained. For electronic integration in Eq. (14), a $30 \times 30 \times 30$ k-point grid ($Fm\bar{3}m$) were employed, and the Dirac deltas were approximated with Gaussian functions with a width of 0.016 Ry ($Fm\bar{3}m$).

# IV. RESULTS AND DISCUSSION

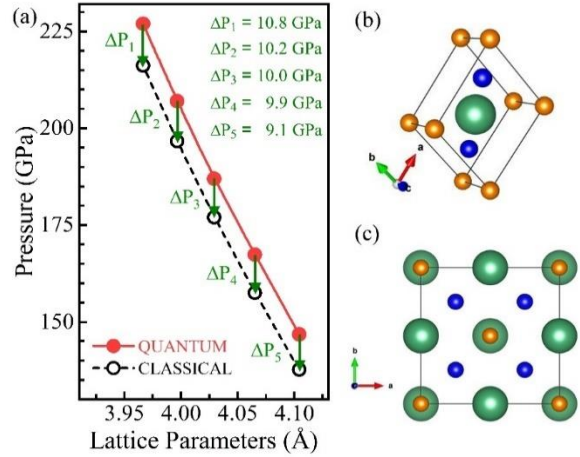

FIG. 1. Comparison between the classical (black dotted line) and quantum (red solid line) pressures as a function of the lattice parameter (a) for $Fm\bar{3}m$-NbH$_3$. The classical pressure is obtained from the Born-Oppenheimer energy surface (BOES), and the quantum pressure is obtained from the SSCHA free energy. Here are shown the differences at five pressures for the same lattice parameter. The crystal structure of $Fm\bar{3}m$-NbH$_3$ is shown with primitive cell (b) and unit cell (c), where the blue and yellow small spheres represent H atoms, the green big spheres represent Nb atoms, respectively.

Recent experiments[49] on niobium poly-hydride at 187 GPa observed a superconducting critical temperature ($T_c$) of 42 K, and suggested that the crystal is NbH$_3$ in $Fm\bar{3}m$ phase based on synchrotron radiation. In this work, we focus mainly on this structure and apply a wide range of pressures to explore the impact of quantum ionic fluctuations on its structural and vibrational properties. The $Fm\bar{3}m$ phase NbH$_3$ has the face-centered symmetry, as illustrated in Fig. 1(b) (primitive cell) and Fig. 1(c) (surface (001) of the cubic unit). In the conventional cubic unit, which contains 4 primitive cells, Nb atoms locate at the body center and 12 edge centers, contributing 4 atoms to the unit. H atoms has two distinct types of locations. The first type, denoted by H1(H2), contains the 8 midpoints between the vertex and the body center, forming interstitial sites. The second type is constituted by the 8 vertices and 6 face centers of the cube, contributing 4 atoms to the unit, denoted as H3. As the SSCHA enforces symmetry, the internal coordinates of the structure remain unaffected by quantum effects.

Nonetheless, the cubic lattice parameter is sensitive to quantum effects. Figure 1(a) reveals that incorporating ionic quantum corrections significantly alters the prediction of pressure. At a given lattice parameter, the classical calculation (yielding harmonic pressure or classical pressure) consistently underestimates the pressure by approximately 10 GPa compared to the pressure obtained using the SSCHA method (referred to as anharmonic pressure or quantum pressure). Similar corrections at the level of 10 GPa have been observed in hydrogen-rich alloys like H$_3$S[33], LaH$_{10}$[39], AlH$_3$[41], and ScH$_6$[42]. The pressure 187 GPa used in the experiment stabilized a structure which has a pressure of only 177 GPa when evaluated with classical calculations. The lattice parameter reported in experiment is 4.09 Å at 184 GPa[49], which is 1.5% larger than the SSCHA prediction at 0 K. However, this discrepancy increases to 2% when compared to harmonic calculations. The hydrogen-hydrogen distance (H~H) is approximately 1.745 Å in the anharmonic case and 1.739 Å in the harmonic case, while experimental measurements[49] yield a value of about 1.77 Å, which is also 1.5% and 2% larger than the calculated values at different levels.

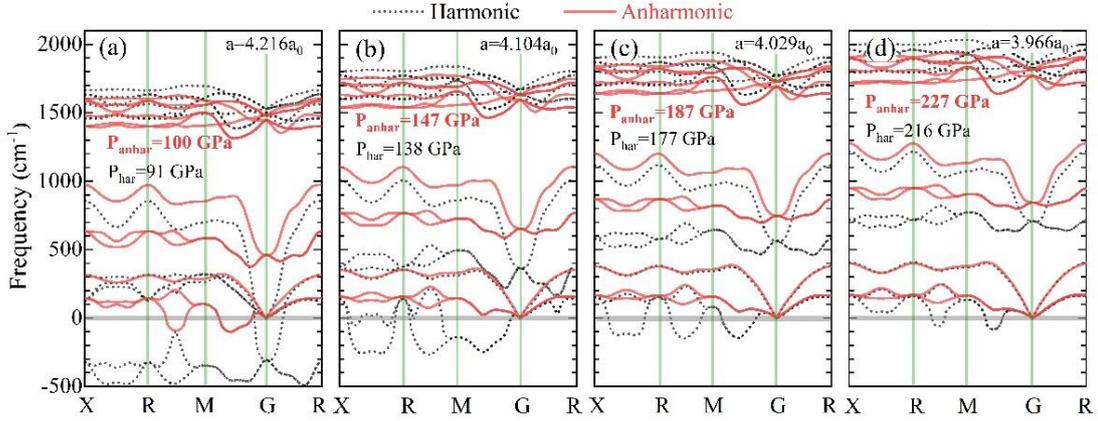

FIG. 2. Comparison between the harmonic (black dotted lines) and anharmonic (red solid lines) phonon spectra of the cubic high-symmetry $Fm\bar{3}m$ phase at different lattice parameters: (a) $4.216a_0$, (b) $4.104a_0$, (c) $4.029a_0$, (d) $3.966a_0$. The anharmonic spectra are obtained from $\boldsymbol{D}^{(F)}$ and correspond to the static limit of the SSCHA dynamical theory. The pressure calculated classically (harmonic calculation) and with quantum effects (anharmonic calculation) is marked in each case. The regions of positive and negative frequencies, the latter of which represent actually imaginary frequencies, are separated with a grey solid line.

As shown in Fig. 2, the quantum and anharmonic effect leads to significant corrections in the phonon spectrum. With these effects included, phonon frequencies of the low-energy acoustic branches and mid-energy optical ones are roughly promoted in several regions of the Brillouin zone (BZ), while the high-energy optical phonons are softened. Similar results of high-energy optical modes have also been reported for the $Pm\bar{3}n$ phase of $AlH_3$[41] and $Pm\bar{3}$ phase of $AlM(M = Hf, Zr)H_6$[61]. With the anharmonic corrections, the high-energy optical modes, mid-energy optical modes and low-energy acoustic modes are well separated, while without them, the latter two get mixed at low pressures (Fig. 2 (a) and (b)).

Very importantly, Fig. 2 implies that the quantum and anharmonic effects can play a vital role to stabilize $Fm\bar{3}m$ phase $NbH_3$ at low pressures. As shown in Fig. 2(b) and Fig. 2(c), the anharmonic corrections eliminate the imaginary phonon modes in the classic results, indicating the dynamical stability at 147 and 187 GPa, consistent with the experimental identification of $Fm\bar{3}m$ phase at 187 GPa. An interpolation process examining the dependence of phonon frequency on pressure provides an estimate of the minimum pressure required to stabilize the crystal. The classical calculations yield a value of 290 GPa, while including anharmonic corrections reduces this estimate to approximately 145 GPa, representing nearly a 50% decrease. Thus, our calculations indicate that the $Fm\bar{3}m$ phase $NbH_3$ can maintain dynamical stability at pressures as low as approximately 145 GPa, at least 147 GPa. However, it likely that $Fm\bar{3}m$-$NbH_3$ loses its thermodynamic stability below 184 GPa, making lower-hydrogen compositions, such as $NbH_{2.5}$ and $NbH_2$[50]. It might be the reason that synthesis is difficult to achieve and the superconductivity was not observed between 145 GPa and 184 GPa[49].

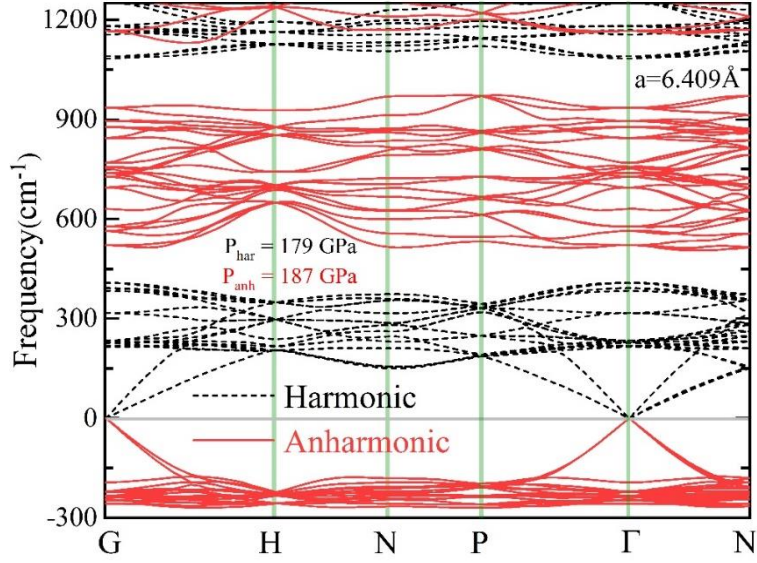

FIG. 3. Comparison between the harmonic (black dashed lines) and anharmonic (red solid lines) phonon spectra of the cubic high-symmetry $I4_3d$ phase at lattice parameters: $6.409a_0$. The region beyond 1250 cm$^{-1}$ is not depicted in the figure. The anharmonic spectra are obtained from $D^{(F)}$ and correspond to the static limit of the SSCHA dynamical theory. The pressure calculated classically (harmonic calculation) and with quantum effects (anharmonic calculation) is marked. The regions of positive and negative frequencies, the latter of which represent imaginary frequencies, are separated with a grey solid line.

Although the $I4_3d$ phase was predicted to be the most thermodynamically and dynamically stable structure in previous classical calculations[51], it was not observed in recent experiments[49]. To investigate this further, we calculated the phonon spectra of the $I4_3d$ phase using the self-consistent harmonic approximation (SSCHA) code. As shown in Fig. 3, the anharmonic phonon modes derived from the Hessian of the free energy ($\mathcal{F}(R_{eq})$) within the SSCHA reveal significant instabilities in all regions of the Brillouin zone at 187 GPa. This finding readily explains the absence of the $I4_3d$ phase in recent high-pressure experiments, eliminating the need to calculate enthalpy of the $Fm\bar{3}m$ and $I4_3d$ structures. Due to computational resource limitations, we did not perform calculations at even higher pressures. The calculations are actually not necessary since the magnitude of imaginary frequencies reach 300 cm$^{-1}$, indicating that significantly higher pressures than those obtained in high-pressure experiments are needed to achieve dynamical stability for the $I4_3d$ structure.

In summary, the phonon spectra reveal that for the dynamical stability of the $Fm\bar{3}m$ and $I4_3d$ structures between 145 GPa to 187 GPa, conclusions from harmonic and anharmonic calculations are entirely opposite. The anharmonic results successfully explain the high-pressure experimental findings[49], indicating that performing anharmonic calculations is essential for an accurate prediction of the vibrational properties of NbH$_3$.

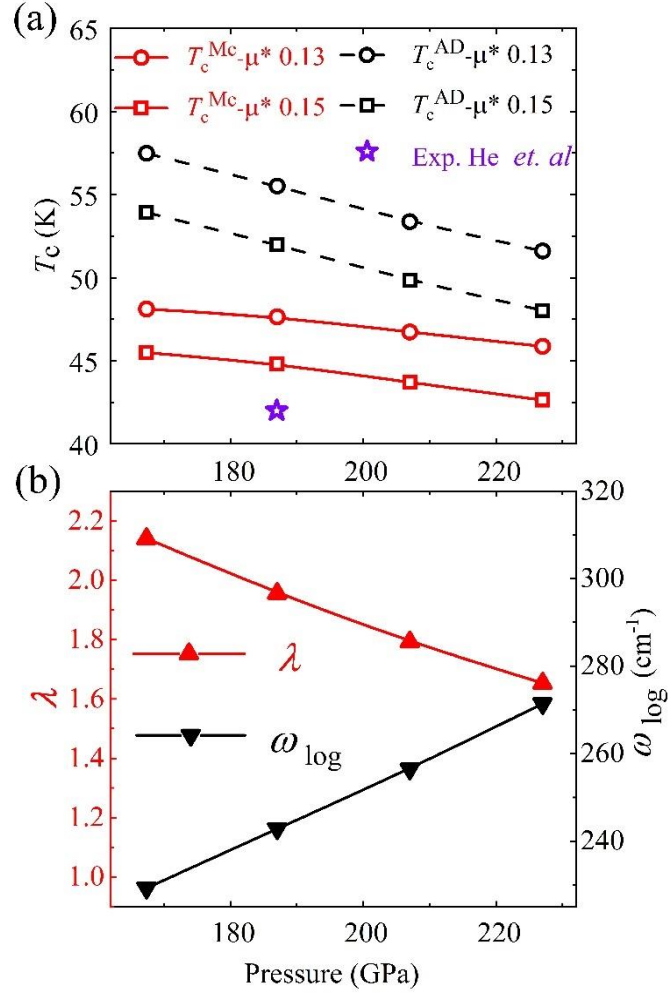

FIG. 4. (a) Superconducting critical temperature $T_c$ evaluated with the McMillan equation (red solid lines) and Allen-Dynes modified McMillan equation (black dashed line) as a function of pressure considering anharmonic effects for $Fm\bar{3}m$-NbH$_3$. $T_c$ calculated with $\mu^*$ =0.13 and 0.15 is plotted with circle symbols and inverted triangle symbols, respectively. The $T_c$ measured in the experiment by He *et. al*[49] is also included with violet star symbol. (b) Electron phonon coupling constant $\lambda$ (red solid line with star symbols) and the average logarithmic frequency $\omega_{\log}$ (black solid line with square symbols) as a function of pressure considering anharmonic effects for $Fm\bar{3}m$-NbH$_3$.

The electron-phonon coupling properties and the resulting superconducting critical temperature ($T_c$) are calculated within the framework of the SSCHA and conventional superconducting theory. As shown in Fig. 4(a), $T_c$ from both the McMillan equation and Allen-Dynes modified McMillan equation decreases monotonously with pressure over the range of 167 to 227 GPa, with the Coulomb pseudopotential $\mu^*$ set to typical values of 0.13 and 0.15. $T_c$ from the Allen-Dynes equation decreases more rapidly, dropping by about 5 K over the 167-227 GPa range, compared to $T_c$ from the McMillan equation, which shows a total drop of less than 2 K. The suppression of $T_c$ by pressure is mainly resulted from the overall hardening of the optical phonon modes imposed by compression, which can be seen in Fig. 2. Raised phonon frequencies with pressure causes the decrease of the electron-phonon coupling (EPC) constant $\lambda$ (eq. (7)), resulting in the reduction of $T_c$ (eq. (5) and (6)). $T_c$ from the McMillan equation with $\mu^*$ set to 0.15 at 187 GPa is 44 K, close to the measured value of 42 K in recent experiment[49]. The EPC constant $\lambda$ declines from 2.14 to 1.65 across the pressure range, while the average logarithmic frequency $\omega_{\log}$ increases from 230 to 271 cm$^{-1}$, as shown in Fig. 4(b). This together results in a relatively mild variation in $T_c$ with pressure. Given the large value of λ, estimating $T_c$ using the McMillan and Allen-Dynes formulas may be inaccurate. Therefore, we solved the Migdal-Eliashberg equation based on the calculated Eliashberg function $\alpha^2F(\omega)$ with μ set to 0.15. The results show a significant increase in $T_c$ compared to the above estimates, reaching around 67 K at 187 GPa, which is more than 20 K higher than the experimental value.

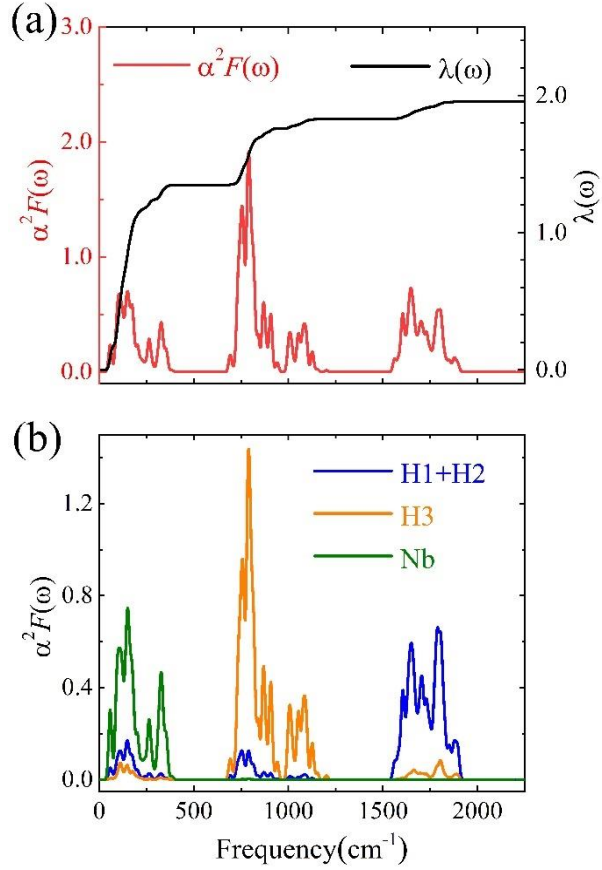

FIG. 5. (a) Anharmonic spectral function $\alpha^2F(\omega)$ (red solid lines) and integrated electron-phonon coupling constant $\lambda(\omega)$ (black solid lines) at 187 GPa (quantum pressure). (b) The projected $\alpha^2F(\omega)$ onto Nb (green), H1 and H2 (blue) and H3 (yellow) at anharmonic levels calculated with lattice parameter 4.029$a_0$, which corresponds in the quantum anharmonic case at 187 GPa.

For further exploration, the Eliashberg spectral function $\alpha^2F(\omega)$ and its integral $\lambda(\omega)$ for anharmonic scenario at 187 GPa are displayed in Fig. 5(a). And the contributions from each specific atom to $\alpha^2F(\omega)$, obtained by counting only the contribution from the atom itself in $\alpha^2F(\omega) = \sum_{\bar{a}\bar{b}} \alpha^2 F_{\bar{a}\bar{b}}(\omega)$ [$\alpha^2 F_{\bar{a}\bar{b}}(\omega)$ can be trivially obtained from Eqs. (14)], are presented in Fig. 5(b). Moreover, the phonon density of states (PDOS) and its atomic projections are also shown in Fig. 6. The contributions from each type of atoms to both $\alpha^2F(\omega)$ and PDOS are distinctly separated. From the figures, three conclusions can be drawn. (i) The low-energy acoustic modes, with frequencies below 400 cm$^{-1}$ and primarily originating from the heavy Nb atoms, contribute approximately 1.36 to the EPC constant λ, accounting for about 70% of its total value of 1.96. (ii) The H3 atoms, forming a face-centered cubic lattice around each Nb atom as its center, generate the mid-energy optical phonons with frequencies between 700 and 1200 cm$^{-1}$, contributing around 0.4 to λ. (iii) The high-energy optical modes, with frequencies above 1500 cm$^{-1}$, originate from the H1(H2) atoms located at interstitial sites. Compared to the H3 atoms, these H1(H2) atoms are more tightly bound by the central Nb atom, resulting in higher vibrational frequencies. These high-energy optical phonons contribute about 0.2 to λ. Combining Figs. 5(b) and 6, one can easily note that, despite a lower phonon density of states, the H3 atoms contribute more to λ than the H1(H2) atoms. This can be readily comprehended by viewing eq. (7), where lower-frequency phonons with larger $α^2F(ω)$ contribute more to the λ integral. The conclusion that the heavy atoms (Nb), rather than the light H atoms, contribute the majority to λ is in stark contrast to some other hydrogen-rich superconductors like $H_3S$[2-26]. This difference likely explains the relatively low $T_c$ of around 40 K observed here.

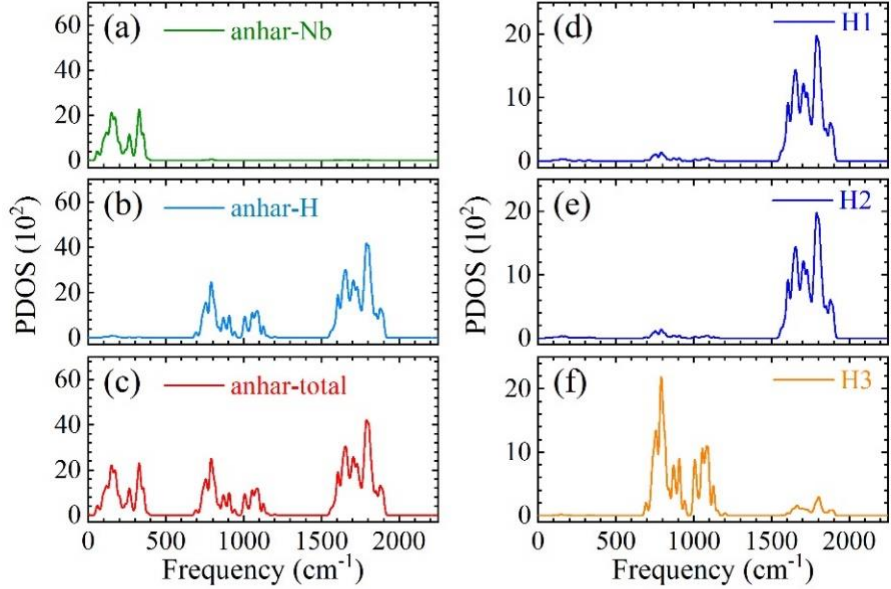

FIG. 6. Total phonon density of states (PDOS) is shown in (c), with contributions from Nb atoms in (a), all H atoms in (b), H1 and H2 atoms in (d) and (e), and H3 atoms in (f) at 187 GPa.

| | **Lattice parameter: 3.859Å** | | | | | | | |
|---|---|---|---|---|---|---|---|---|
| | Pressure | $T_C^{AD}$ (K) | | $T_C^{MC}$ (K) | | $T_C^{ME}$ (K) | | λ | $ω_{\log}$ (cm$^{-1}$) |
| | | 0.13 | 0.15 | 0.13 | 0.15 | 0.13 | 0.15 | | |
| ***Anharmonic*** | 310 GPa | 35.5 | 31.7 | 33.1 | 29.7 | 52 | 49 | 1.03 | 364.1 |
| ***Harmonic*** | 299 GPa | 51.3 | 47.9 | 43.4 | 41.2 | 64 | 60 | 1.83 | 240.9 |

TABLE I. Superconducting critical temperature $T_c$ with $\mu^*$ =0.13 and 0.15 evaluated by the McMillan equation ($T_c^{MC}$), Allen-Dynes modified McMillan equation ($T_c^{AD}$) and Migdal-Elishberg equation ($T_c^{ME}$) compared under considering anharmonic effects (Anharmonic) and harmonic approximation (Harmonic) for $Fm\bar{3}m$-NbH$_3$ with lattice parameter: 3.859Å. The pressure, electron phonon coupling constant $\lambda$ and the average logarithmic frequency $\omega_{\log}$ are also compared with the same lattice parameter.

Previous studies have highlighted that the anharmonic corrections to the phonon spectra may significantly affect the computed $T_c$[31, 34, 41]. However, comparing $T_c$ at harmonic and anharmonic levels for $Fm\bar{3}m$-NbH$_3$ at experimentally accessible pressures is challenging due to the presence of harmonic imaginary phonons below 290 GPa. Therefore, we examine the effects of anharmonicity on $T_c$ at a higher pressure of 310 GPa, corresponding to a harmonic pressure of 299 GPa, where the dynamical stability is achieved in the harmonic calculation. $T_c$s evaluated from the McMillan equation, Allen-Dynes equation and Migdal-Elishberg equation using both the harmonic and anharmonic phonons are presented in Tab. 1. It is obvious that the harmonic approximation overestimates $T_c$. As shown in Tab. 1. using the McMillan equation with typical values of 0.13 and 0.15 for $\mu^*$, the harmonic approximation yields $T_c$ values of 43 K and 41 K, while considering anharmonic quantum effects reduces these values to 33 K and 30 K, respectively. This discrepancy of approximately 10 K amounts to a 30% overestimation in the harmonic case. At this pressure, the EPC constant $\lambda$ from harmonic and anharmonic calculations are 1.83 and 1.03, respectively, representing a significant difference of 80%. Thus, neglecting the quantum and anharmonic nature of ions in $Fm\bar{3}m$-NbH$_3$ can lead to a significant overestimation of the EPC constant $\lambda$ and the resulting $T_c$.

## V. CONCLUSIONS

In summary, this study provides a detailed first-principles analysis of the role of quantum ionic fluctuations and anharmonicity in stabilizing the high-pressure $Fm\bar{3}m$ phase of NbH$_3$ and influencing its superconducting properties. Using the stochastic self-consistent harmonic approximation to capture anharmonic effects beyond standard perturbation theory, we find that the $Fm\bar{3}m$ phase remains dynamically stable down to at least 145 GPa—significantly lower than the harmonic predictions. Recent experiments at 187 GPa identified $Fm\bar{3}m$ NbH$_3$ as the likely crystal structure, while the $I4_3d$ phase predicted by earlier harmonic calculations as the most favorable structure is not observed. Moreover, the anharmonic calculations in this work confirm the dynamical instability of the $I4_3d$ phase at 187 GPa, consistent with the experimental observations. The electron-phonon coupling properties and resulting superconducting critical temperature ($T_c$) of $Fm\bar{3}m$ NbH$_3$ are calculated using the McMillan, Allen-Dynes modified McMillan and Migdal-Elishberg equations, with the Coulomb pseudopotential $\mu^*$ set to typical values of 0.13 and 0.15. At 187 GPa, the calculated $T_c$ from McMillan equation is 44 K for $\mu^*$ = 0.15, closely matching the experimental result. Further analysis of the Eliashberg spectral function $\alpha^2F(\omega)$ and phonon density of states reveals that, the heavy Nb atoms contribute roughly 70% to the electron-phonon coupling constant $\lambda$ and thus play a decisive role in determining $T_c$, while hydrogen atoms have a much smaller influence on $\lambda$ and $T_c$. This contrasts with some other hydrogen-rich superconductors and may explain the relatively modest $T_c$ of around 40 K observed here. The effects of the quantum and anharmonic fluctuations of ions on $T_c$ are investigated at a high pressure of 310 GPa, corresponding to a harmonic pressure of 299 GPa, where the dynamical stability is achieved in the harmonic calculation. Calculations at both the harmonic and anharmonic levels suggest that, the harmonic approximation, which neglects the quantum and anharmonic effects, significantly overestimates λ and the resulting $T_c$. Our findings highlight the critical role of quantum anharmonic effects in determining the dynamical stability of the $Fm\bar{3}m$ and $I4_3d$ phases and the superconducting properties of the $Fm\bar{3}m$ phase.

## ACKNOWLEDGMENTS

This research was supported by the Natural Science Foundation Project (Grant No. 20230101280JC) of Jilin Provincial Department of Science and Technology.